\documentstyle[12pt,epsfig,amsmath,amssymb]{article}
\newcommand{\f}[2]{\frac{#1}{#2}}
\newcommand{\la}{\langle}
\newcommand{\ra}{\rangle}

\newcommand{\Oc}{{\cal O}}
\newcommand{\de}{\partial}

\newcommand{\eqp}{\stackrel{P}{=}}

\begin{document}
\rightline{IFUP--TH}

\vskip 1cm

\centerline{{\bf Behavior near $\theta=\pi$ of the mass gap in the 2D O(3) non--linear sigma model}}
\vskip 5mm
\centerline{B.~All\'es$^{\rm a}$, M. Giordano$^{\rm b}$, A. Papa$^{\rm c}$}

\centerline{\it $^{\rm a}$INFN Sezione di Pisa, Largo Pontecorvo 3, 56127--Pisa, Italy}

\centerline{\it $^{\rm b}$Institute for Nuclear Research of the Hungarian}
\centerline{\it Academy of Sciences (ATOMKI), Bem t\'er 18/c, H--4026 Debrecen, Hungary}

\centerline{\it $^{\rm c}$Dipartimento di Fisica, Universit\`a della Calabria and}
\centerline{\it  INFN Gruppo Collegato di Cosenza, Arcavacata di Rende, 87036--Cosenza, Italy}

\begin{abstract}
The validity of the Haldane's conjecture entails that the mass gap of the 2--dimensional O(3) non--linear sigma model with a
$\theta$--term must tend to zero as $\theta$ approaches the value $\pi$
by following a precise law. In the present
paper we extract the related critical exponents by simulating the model at imaginary $\theta$.
\end{abstract}

\vfill\eject


\section{Introduction}
\label{sec0}

The proposal of Haldane in Refs.~\cite{haldane1,haldane2,haldane3},
regarding the absence of a gap in the spectrum of 1D (throughout the
paper $n$D means $n$--dimensional) quantum chains of half--integer
spins interacting through an antiferromagnetic coupling, prompted a
great deal of work aimed at checking its correctness. Apart from
numerical simulations and direct analytical scrutiny on such 1D
quantum
chains~\cite{lieb,botet,affleck1,affleck2,affleck3,schollwock}, an   
important step towards the clarification of the validity of this
proposal was achieved by demonstrating that such chains and the 2D
O(3)--invariant non--linear sigma model with a topological term at
vacuum angle $\theta=\pi$ share the same long--distance behavior, see 
Refs.~\cite{haldane1,haldane2,Affleck:1991tj,fradkin}. 
Hence, the above property can also be verified by studying the 2D O(3) 
non--linear sigma model near $\theta=\pi$. Although direct Monte Carlo
simulations are currently unfeasible for $\theta\not=0$ due to the
sign problem\footnote{Nevertheless, an effort has recently been 
pursued to avoid the sign problem by simulating the model at
$\theta\not=0$ after demonstrating the equivalence of its 
continuum limit with that of the dual of the SU(2) principal chiral
model with a fixed radial part, see Ref.~\cite{torrero}.} 
(see (\ref{latticeaction}) below), several tricks have been contrived
to analyze the model at non--zero $\theta$ and, particularly, at
$\theta=\pi$. They are: {\it (i)} in Refs.~\cite{bietenholz,BNPW,dFPW}
the distribution of the topological charge was determined at
$\theta=0$ and  then used to reweight the partition function at
$\theta\not=0$; {\it (ii)} in Ref.~\cite{alles1} the mass gap was
extracted as a function of imaginary $\theta$ (with which the sign
problem disappears, see (\ref{latticeaction}) below) and the results 
extrapolated to real $\theta$; {\it (iii)} in Refs.~\cite{azcoiti1,zgz}
a similar method was employed, measuring the expectation value of the
topological charge at imaginary $\theta$, after which a controlled way 
to perform the extrapolation allowed the authors to reduce the
uncertainties. In all cases a decisive confirmation of the Haldane's
conjecture, namely, the mass gap of the 2D O(3) non--linear sigma
model vanishes at $\theta=\pi$, was obtained.  

The equivalence between 1D antiferromagnetic chains of spins and the
2D O(3) non--linear sigma model with $\theta=\pi$ has been further
investigated. It was argued in Refs.~\cite{affleck4,affleck5} that the
critical theory for the half--integer quantum antiferromagnetic spin
chains is the Wess--Zumino--Novikov--Witten (WZNW) model with a
topological coupling $k=1$, defined in
Refs.~\cite{wess,novikov,witten}. This model is the stable fixed point 
of the 2D O(3) non--linear sigma model with a vacuum angle
$\theta=\pi$. The renormalization group considerations of 
Refs.~\cite{affleck4,affleck5} on the WZNW model lead to the
conclusion that the mass gap of the 2D O(3) non--linear sigma model
tends to zero while approaching $\theta=\pi$ from below as
\begin{equation}
  m(\theta)\propto
  (\pi-\theta)^{\epsilon_{_{\rm WZNW}}}
  \left(\log\frac{1}{\pi-\theta}\right)^{-\beta_{_{\rm WZNW}}}\;,  
\label{massWZNW}
\end{equation}
for $0<\pi-\theta\ll1$. The WZNW predictions are $\epsilon_{_{\rm
    WZNW}}\equiv\frac{2}{3}$ and $\beta_{_{\rm WZNW}}\equiv\frac{1}{2}$. 
Therefore, another type of useful check of the Haldane's conjecture 
consists in finding the critical exponents in (\ref{massWZNW}) from
numerical simulations of the 2D O(3) non--linear sigma model with a
non--zero $\theta$--term. In Refs.~\cite{bietenholz,BNPW,dFPW} the
authors compared the numerical results with the theoretical prediction
for the step scaling function, finding good agreement. In this paper
we want to approach this issue in a different way, attempting instead
at a determination of both the critical exponent and the exponent of
the logarithmic correction, from Monte Carlo simulations at imaginary
$\theta$ using the method of Ref.~\cite{alles1} together with the 
improvement procedure of Refs.~\cite{azcoiti1,zgz}.\footnote{As it was 
  apparent in Ref.~\cite{alles1}, the results of the simulations for
  the mass gap alone are too noisy to allow a reasonably clear
  determination of the exponents in (\ref{massWZNW}).} 
A similar approach was used in Ref.~\cite{zgz}, where however the
theoretical expectation for the logarithmic term was used as an input
in the analysis. As we shall see along the present paper, a direct
detection of the power of the logarithmic correction in (\ref{massWZNW}) 
requires an extremely accurate control of the statistics and error
bars, an endeavor that seems to lie beyond present--day
capabilities. It is however possible to bypass this difficulty, by 
combining the analyses of the mass gap and of the topological charge.
The purpose of our paper consists precisely in employing this combined
analysis to retrieve the exponents $\epsilon_{_{\rm WZNW}}$ and
$\beta_{_{\rm WZNW}}$ in (\ref{massWZNW}). 

In Section \ref{sec1} the 2D O(3) non--linear sigma model with a
$\theta$--term is introduced and its main properties briefly
enumerated. In Section \ref{sec2} the Monte Carlo method and related
difficulties shall be presented. In Section \ref{sec3} the basics of
the extrapolation method from imaginary to real $\theta$ will be
explained, while the difficulties related to the presence of
logarithmic corrections in (\ref{massWZNW}) are attacked in Sections
\ref{sec4} and \ref{sec5}. In Section \ref{sec5} also the details of
the data analysis will be spelled out. The conclusions are listed in
Section \ref{sec6}.  

\section{The 2D O(3) non--linear sigma model}
\label{sec1}

The action of the 2D O(3) non--linear sigma model with a
$\theta$--term in the continuum is given by  
\begin{equation}
  \label{continuumaction}
  \begin{aligned}
    S&=A - {\rm i} \theta Q\;, \quad A= \frac{1}{2g}
    \int \hbox{d}^2x \left(\partial_\mu \vec{\phi}(x)\right)^2\;,
    \\
    Q&=\int \hbox{d}^2x \,Q(x)\;,\quad  \\
    Q(x)&\equiv
    \frac{1}{8\pi} \epsilon^{\mu\nu} \epsilon_{abc}
    \phi^a(x) \partial_\mu \phi^b(x)
    \partial_\nu \phi^c(x)\; ,
    \end{aligned}
\end{equation}
where $g$ is the coupling constant, $\theta$ the vacuum angle, $Q(x)$
is the topological charge density and $Q$ the total topological
charge. $\vec{\phi}(x)$ is a 3--component unit vector that represents
a classical spin, the dynamical variable at position $x$. The
renormalized $Q$ takes on integer values because it counts how many
times the spin variables wrap around the unit sphere.

This model enjoys various properties that make it an interesting
object of study in areas ranging from condensed matter to field
theory. In particular, the quantum Hall effect can be studied by it
(see for example Ref.~\cite{pruisken}) and some attributes of field 
theories like asymptotic freedom, spontaneous generation of a gap or
instantonic effects are present in the 2D O(3) non--linear sigma
model, see Ref.~\cite{belavin}. Specifically, the mass gap at
$\theta=0$ has been calculated exactly in Ref.~\cite{hasenfratz}. This 
gap diminishes as $\theta$ increases as shown in
Refs.~\cite{affleck5,controzzi} until reaching zero at $\theta=\pi$ if
(\ref{massWZNW}) holds. 

\section{The Monte Carlo program}
\label{sec2}

We have regularized the model (\ref{continuumaction}) on a square
lattice with periodic boundary conditions by the expression
\begin{equation}
  S_L = A_L -{\rm i} \theta Q_L \;, \quad
  A_L \equiv-\frac{1}{g_L} \sum_{x,\mu} \vec{\phi}(x)
  \cdot \vec{\phi}(x+\widehat{\mu}) \,,
  \label{latticeaction}
\end{equation}
where $Q_L=\sum_xQ_L(x)$ is the total lattice topological charge,
$Q_L(x)$ the lattice topological charge density and $g_L$ is the
bare lattice coupling constant. The standard action $A_L$ used
in~(\ref{latticeaction}) is the simplest one on the lattice that
reproduces $A$ in (\ref{continuumaction}) in the continuum limit.

The topological charge density has been regularized by defining it on
triangles (not on single sites). Every plaquette of a square lattice
can be cut through a diagonal into two triangles. If we call
$\vec{\phi}_1$, $\vec{\phi}_2$ and $\vec{\phi}_3$ the fields at the
sites of the three vertices (numbered counter--clockwise) of one of
these triangles, then the fraction of spherical angle subtended by
these fields is $Q_L(\bigtriangleup)$ which satisfies (see
Ref.~\cite{berg}) 
\begin{equation}
  \exp\left(2\pi {\rm i}  Q_L(\bigtriangleup)\right) =
  \frac{1}{\rho}\Big(
  1+\vec{\phi}_1 \cdot \vec{\phi}_2 +
  \vec{\phi}_2 \cdot \vec{\phi}_3 +
  \vec{\phi}_3 \cdot \vec{\phi}_1 +
  {\rm i} \vec{\phi}_1 \cdot \big(\vec{\phi}_2 \times
  \vec{\phi_3}\big)\Big)\;,
\label{latticeQ2}
\end{equation}
where $\rho^2\equiv 2 (1 + \vec{\phi}_1 \cdot \vec{\phi}_2)
(1 + \vec{\phi}_2 \cdot \vec{\phi}_3)
(1 + \vec{\phi}_3 \cdot \vec{\phi}_1)$ and
$Q_L(\bigtriangleup)\in [-\frac{1}{2},+\frac{1}{2}]$.
Elementary plaquettes can be cut in two ways, but both choices lead to
the same physical results for expectation values. The sum of
$Q_L(\bigtriangleup)$ over all of the triangles yields the so--called
geometric topological charge~$Q_L$, which provides integer values
without requiring a composite operator renormalization. 

A configuration of spins is a set of values of $\vec{\phi}(x)$ for all
lattice points $x$ that yields a definite number if plugged into
expression (\ref{latticeaction}).\footnote{This definition excludes
  the so--called exceptional configurations, to which a value of the
  topological charge cannot be assigned unambiguously, but which
  constitute a set of zero measure~\cite{berg}.}
Monte Carlo simulations permit to collect configurations that are
distributed according to the Boltzmann weight $\exp(-S_L)$ as long as 
$S_L$ is real. Unfortunately, this condition fails to hold in our
problem for $\theta\not=0$. Indeed, the sign problem in the second
term of $S_L$ is evident due to the presence of the imaginary
unit. The existence of this problem makes the model even more 
appealing since similar difficulties appear also in the lattice
regularization of several field theories like QCD at finite baryon
density. To avoid it, we have numerically simulated the action
(\ref{latticeaction}) at imaginary values of $\theta=-{\rm
  i}\vartheta$ ($\vartheta$ is real) and extrapolated the 
results  to real $\theta$. Simulations were done using a Metropolis
algorithm. 

The simulations were all performed at $1/g_L=1.6$ on a square lattice
of lateral size $L=180$. These choices were dictated by the need of
working within a scaling window with as little finite--size and
coarse--graining effects as possible. Specifically, as shown in
Ref.~\cite{alles2}, the size $L=180$ is the one for which the model at
$1/g_L=1.6$ and $\theta=0$ displays a ratio $L/\xi\sim10$ ($\xi$ is 
the correlation length or inverse of the gap). We will see later that,
whereas $\xi$ increases with $\theta$, it decreases for increasing
$\vert\vartheta\vert$. For this reason the ratio $L/\xi$ becomes
larger at non--zero $\vartheta$ and this fact enables us to maintain a
good control on the finite--size effects in every single
simulation. All these features were verified by explicit simulations
on smaller lattice sizes ($L=100$ and $L=60$) obtaining numerically
the same results within errors. 

\begin{figure}[t]
  \centering
  \includegraphics[width=0.9\textwidth]{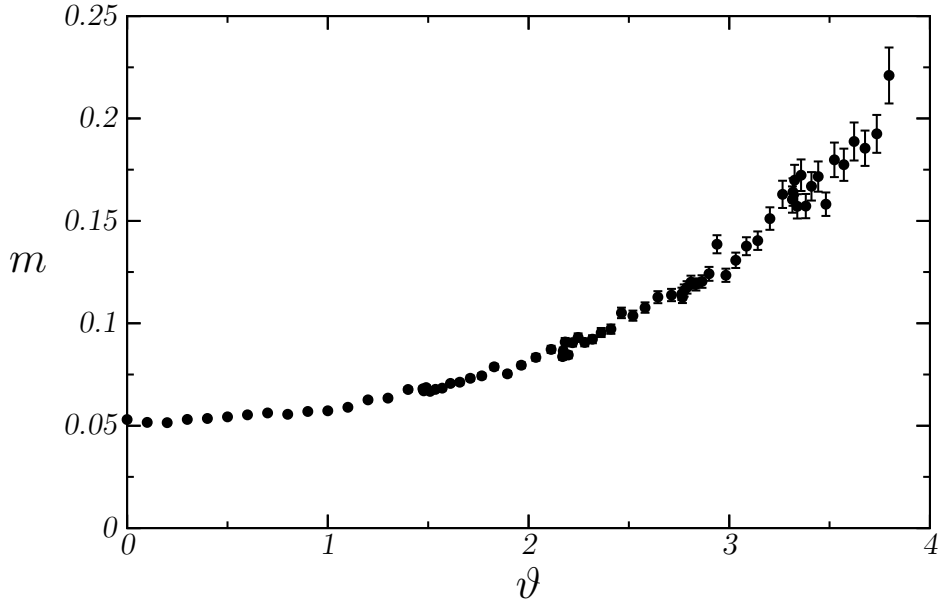}
  \caption{The mass gap $m$, or inverse correlation length, as a
    function of $\vartheta={\rm i}\theta$ for $1/g_L=1.6$.} 
  \label{fig:1}
\end{figure}

For the subsequent analysis, measurements of the topological charge
and of the mass gap are needed. Measurements of the first observable
are obtained by the procedure explained in the text around
(\ref{latticeQ2}) and can be read off during the very Metropolis
steps. To determine the mass gap (the inverse of the correlation
length) we computed the two--point correlation function,
\begin{equation}
  G(x_1,x_2)\equiv\langle\vec{\phi}(0,0)\cdot\vec{\phi}(x_1,x_2)\rangle\;, 
  \label{G}
\end{equation}
where brackets $\langle\cdots\rangle$ indicate the average with the
Boltzmann weight and $x_1$ and $x_2$ are the two components of
$x$. The precise definition of correlation length we employed was 
\begin{equation}
  \xi\equiv\frac{\sqrt{\chi/{\cal F}-1}}{2\sin\pi/L}\;,
  \label{xi2}
\end{equation}
where $\chi$ is the magnetic susceptibility and ${\cal F}$ the
correlation function at the smallest non--zero lattice momentum
$2\pi/L$, 
\begin{equation}
  \label{chiF}
  \begin{aligned}
    \chi&\equiv\sum_{x_1,x_2}G(x_1,x_2)\;,\\
    {\cal F}&\equiv\frac{1}{2}\sum_{x_1,x_2}({\rm e}^{2\pi{\rm i}
      x_1/L}+{\rm e}^{2\pi{\rm i} x_2/L})G(x_1,x_2)\;. 
  \end{aligned}
\end{equation}
Definition (\ref{xi2}) has two advantages. On the one hand the gap
follows from a more straightforward calculation than the one employed
in definitions based on the exponential decay of $G(x_1,x_2)$ (thus
simplifying the error evaluation) and on the other hand the dependence
of $\xi$ on the lattice size $L$ is as negligible as it is for the
above--mentioned exponential decay--based definitions, see
Ref.~\cite{alles3} (thus offering a very robust estimate). Errors were
assessed by blocking. 

We simulated the model for 75 different values of $\vartheta$ spanning
from~0 to 3.7964. For each value of $\vartheta$, 2 million of
thermalized configurations were prepared. Each configuration and the
next one were separated by 100 decorrelation hits and the norms of the
fields ($\Vert\vec{\phi}(x)\Vert=1$ for all $x$) were checked and
reset every 20 Metropolis hits (actually, the whole procedure turned
out to be numerically very stable since the residuals
$\vert\Vert\vec{\phi}\Vert-1\vert$ always remained negligibly small  
and in any case well within the computer accuracy). The Marsaglia
random number generator was utilized. 

\begin{figure}[t]
  \centering
  \includegraphics[width=0.9\textwidth]{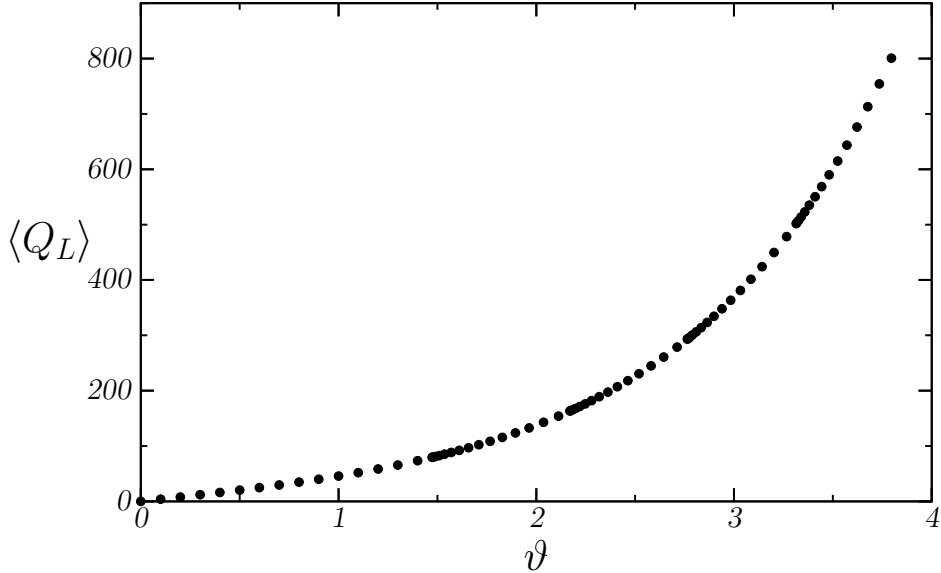}
  \caption{The expectation value of the 
    total topological charge, $\langle Q_L\rangle$, as a function of
    $\vartheta={\rm i}\theta$ for $1/g_L=1.6$. Error bars are smaller
    than the size of the symbols.
  }
  \label{fig:2}
\end{figure}

The numerical results for the mass gap $m=1/\xi$ and the average
topological charge are shown in Figs.~\ref{fig:1} and \ref{fig:2}, as
functions of $\vartheta={\rm i}\theta$. Since we want to investigate
the behavior at real $\theta\simeq\pi$, we need to perform the
analytic continuation of our numerical data, which is known to be a
difficult problem. This issue is discussed in the next Section.

\section{The method of scaling transformations}
\label{sec3}

The basic technique we want to use in order to understand the critical
behavior at $\theta=\pi$ is that of scaling transformations proposed
in Ref.~\cite{ADGV1} and subsequently used in
Refs.~\cite{ADGV2,azcoiti1,AFV,zgz}. This technique provides a 
controllable way to perform the analytic continuation of results
obtained simulating at imaginary $\theta$. We give here a brief
description of this approach; more details can be found in the
above--mentioned references.  

Originally, the approach was proposed in order to study the behavior of
the expectation value of the  topological charge for systems with a
$\theta$--term near $\theta=\pi$. Instead of working with the
topological charge itself, it turns out to be more convenient to use
the quantity 
\begin{equation}
  \label{eq:method1}
  y(z)=\f{\la Q_L \ra}{V\tanh\f{\vartheta}{2}} \equiv
  \f{q_\vartheta}{\tanh\f{\vartheta}{2}}\,,\qquad 
  z=\cosh\f{\vartheta}{2}\,,\quad z\ge 1\,,
\end{equation}
where $q_\vartheta$ denotes the vacuum expectation value of $Q_L$ over
the volume $V$ for imaginary $\theta=-{\rm i}\vartheta$, making
explicit by means of a subscript its dependence on $\vartheta$ for
future convenience. Upon analytic continuation back to real values,
$\vartheta\to {\rm i}\theta$, one has 
\begin{equation}
  \label{eq:method1bis}
  y(z) =-{\rm i}\f{q_{{\rm i}\theta}}{\tan\f{\theta}{2}}
  \,,   \qquad  z=\cos\f{\theta}{2}\,,\quad z\le 1\,,
\end{equation}
{\it i.e.}, in terms of $z$ the analytic continuation is simply an
extrapolation from $z\ge 1$ to $z\le 1$.\footnote{Notice that $y(z)$
  remains real also for $z\le 1$.} The next step consists in
performing this extrapolation not by using directly $y(z)$ as a
function of $z$, but rather by relating $y(z)$ to $y_\lambda(z)\equiv
y({\rm e}^{\f{\lambda}{2}}z)$, {\it i.e.}, by trying to determine the
function $y_\lambda(y)$. The assumption usually made is that $y(z)$ is
a monotonically increasing function of $z$, and that moreover it
vanishes only for $z=0$ ({\it i.e.}, $\theta=\pi$), which in physical
terms corresponds to the absence of phase transitions sending the
topological charge to zero in the interval
$\theta\in(0,\pi)$. Actually, this is indeed the case for the models
where the exact solution is known. The quantity $y_\lambda$ is then a
monotonic function $y_\lambda(y)$ of $y$, with the property that 
$y_\lambda=0$ at $y=0$. The expectation is that $y_\lambda(y)$ is a
smooth function, so that starting from the smallest values of $y$ that
can be obtained by numerical simulations at real $\vartheta$, one can  
reliably extrapolate towards $y=y_\lambda=0$, {\it i.e.}, in the
region corresponding to real $\theta=-{\rm i}\vartheta$. From this 
point of view, for asymptotically free systems the situation gets more
and more favorable as one gets closer to the continuum limit, since
topological fluctuations become suppressed. 

In the 2D O(3) non--linear sigma model, the behavior of the total topological
charge near $\theta=\pi$ is related to that of the mass
gap.\footnote{The well known problems that appear when taking the
  continuum limit of topological observables are not relevant for the
  problem at   hand, see the discussion in Ref.~\cite{zgz} based on
  the results of Refs.~\cite{BDSL,Nogradi}.} According to Haldane's 
conjecture, the mass gap vanishes as $m(\theta) \sim (\pi
-\theta)^\epsilon \sim z^\epsilon$ (up to logarithmic corrections) for
$\theta\to\pi$ from below, see Eq.~\eqref{massWZNW}, and since 
$q_{{\rm i}\theta} \propto \de m^2/\de\theta$, one expects $y\sim (\pi
 -\theta)^{2\epsilon} \sim z^{2\epsilon}$. One can then determine the
 critical exponent $\epsilon$ of the mass gap by extrapolating the
 following effective exponent, 
\begin{equation}
  \label{eq:gammal}
  2\epsilon_q(y) \equiv \f{2}{\lambda}\log\f{y_\lambda(y)}{y}\,,
\end{equation}
towards $y=0$, {\it i.e.}, towards $\theta=\pi$. One easily sees that
$\epsilon_q(0)=\epsilon$. In principle, the same kind of technique can
be used to study the behavior of any observable near $\theta=\pi$, and
in particular one can work directly with the mass gap. Defining
$m_\lambda(z) = m({\rm e}^{\f{\lambda}{2}}z)$, re--expressing it as
$m_\lambda(y)$, and defining the effective exponent
\begin{equation}
  \label{eq:gammal_m}
  \epsilon_m(y) \equiv \f{2}{\lambda}\log\f{m_\lambda(y)}{m(y)}\,,
\end{equation}
one finds again that $\epsilon_m(0)=\epsilon$.\footnote{We assume here 
that there are no phase transitions for $\theta\in(0,\pi)$ that send
the mass to zero.} 

Despite the successful application of the method described above to
several models~\cite{ADGV1,ADGV2,AFV}, it turns out that the direct
application of Eqs.~\eqref{eq:gammal} and \eqref{eq:gammal_m} to the
analysis of numerical data in the 2D O(3) non--linear sigma model is
hampered by the presence of logarithmic corrections to the critical
behavior shown in (\ref{massWZNW}), see Ref.~\cite{zgz}. In the next
Section we briefly discuss the form of these logarithmic corrections,
and in Section \ref{sec5} we propose a method to overcome the related
difficulties.

\section{Critical behavior of the 2D O(3) non--linear sigma model
  with a topological term}
\label{sec4}

The appropriate WZNW model describing the critical behavior 
of the 2D O(3) non--linear sigma model with topological term near
$\theta=\pi$ has been studied in Refs.~\cite{affleck4,affleck5,controzzi}. In
particular, in Ref.~\cite{affleck5} the authors have determined the
relation between the correlation length $\xi$ and the coupling $\tilde
g\sim \pi - \theta$ of the relevant perturbation near $\theta =\pi$, 
which reads 
\begin{equation}
  \label{eq:1}
  \f{1}{\tilde g} \propto \xi^{\f{3}{2}}(\log\xi)^{-\f{3}{4}}[1 +
  \Oc(1/\log\xi) ]\,. 
\end{equation}
Instead of working out the corresponding prediction for the critical
behavior of the correlation length, we will use the more general
expression 
\begin{equation}
  \label{eq:2}
   \f{1}{\tilde g} = K \xi^a (\log\xi)^{-b} u(\log\xi)\,,
\end{equation}
with $K$ some constant and some function $u(x)=1 +\sum^\infty_{k=1}u_k
x^{-k}$, which reduces to the results of Ref.~\cite{affleck5} for
$a=\f{3}{2}$ and $b=\f{3}{4}$ (and with an appropriate $u$). 
The reason why we do the calculation in this generalized setting is
that we want a general expression for a vanishing mass gap, not
relying on the details of the relevant critical model, which can be
guessed on general grounds, and which can be used in principle to
determine the critical exponents from numerical data, without knowing
in advance the values of $a$ and $b$. This is different from the
approach of Refs.~\cite{bietenholz,BNPW,dFPW} where the theoretical
expectation for the critical behavior was used as an input of the
numerical analysis. Since the theoretical prediction is strictly valid
only in the continuum limit, a shortcoming of this approach is that it
cannot be used to map the full phase diagram of the $O(3)$ non--linear
sigma model at $\theta=\pi$ as the coupling is varied.\footnote{For
  example, at strong coupling the system is expected to undergo a
  first--order phase transition at $\theta=\pi$.} On the other hand,
our approach is sufficiently general and could be applied to the study
of this problem. 

We now derive the critical behavior of the correlation
length. Eq.~\eqref{eq:2} can be inverted by solving for $\xi$  
iteratively. The solution has the form
\begin{equation}
  \label{eq:5}
  \log\xi = \f{1}{a}\log\f{1}{\tilde g K} + 
\f{b}{a}\log\log\f{1}{\tilde g K} +
\f{b}{a}\log\f{1}{a} +
\sum_{l=1}^\infty \sum_{j=0}^l C^j_l
\f{\left(\log\log\f{1}{\tilde g K}\right)^j}{\left(\log\f{1}{\tilde g
      K}\right)^l}\,, 
\end{equation}
where $C^j_l$ are constants. For our purposes we shall use the
variable $z = \cos\f{\theta}{2}$, which behaves as $z\simeq
(\pi-\theta)/2$ near $\pi$ and is therefore proportional to $\tilde
g$. Subleading terms in the expansion of $z$ are powers in $\tilde g$
(and vice versa) and so will be discarded, since we are considering 
here only logarithmic terms, which dominate the critical behavior. 
We obtain
\begin{equation}
  \label{eq:6}
  \log\xi \eqp \f{1}{a}\log\f{1}{z} + 
  \f{b}{a}\log\log\f{1}{z} +
  \bar C^0_0 +
  \sum_{l=1}^\infty \sum_{j=0}^l \bar C^j_l
  \f{\left(\log\log\f{1}{z}\right)^j}{\left(\log\f{1}{z}\right)^l}\,, 
\end{equation}
where the mark $P$ over the equals sign indicates that the equality
holds up to terms which are proportional to powers of $z$, and $\bar
C^j_l$ are constants. Recalling now that the mass gap is $m=1/\xi$,
and exponentiating Eq.~\eqref{eq:6}, we finally get
\begin{equation}
  \label{eq:7}
  m \eqp z^\epsilon \left(\log\f{1}{z}\right)^{-\beta} \exp\left\{
    -\sum_{l=0}^\infty \sum_{j=0}^l \bar C^j_l
    \f{\left(\log\log\f{1}{z}\right)^j}{\left(\log\f{1}{z}\right)^l}
  \right\}\,,
\end{equation}
where $\epsilon=\f{1}{a}$ and $\beta = \f{b}{a}$. Substituting the
values appearing in Eq.~\eqref{eq:1}, one obtains the theoretical
expectation for the critical exponents, 
$\epsilon_{_{\rm WZNW}}=\f{2}{3}$ and $\beta_{_{\rm WZNW}}=\f{1}{2}$. 

Even though most of the coefficients in Eq.~\eqref{eq:7} are not fully 
determined, as the detailed form of the function $u$ in
Eq.~\eqref{eq:2} is largely unknown, nevertheless the coefficients 
$\bar C^l_l=C^l_l$, $l\ge 1$ can be determined exactly, as they do not
depend on $u$, and the corresponding terms can be resummed. Setting
$w=\log\f{1}{z}$,  $m_0 = {\rm e}^{-\bar C^0_0}$, and 
\begin{equation}
  \label{eq:9}
  \bar u(w) = \exp\left\{
    -\sum_{l=1}^\infty \sum_{j=0}^{l-1} \bar C^j_l
    \f{\left(\log w\right)^j}{w^l}
  \right\} = 1 + \Oc(1/w)\,,
\end{equation}
we finally obtain
\begin{equation}
  \label{eq:10}
  m \eqp m_0 {\rm e}^{-\epsilon w}w^{-\beta}
  \left[1 + \f{\beta}{\epsilon}\f{\log w}{w}
  \right]^{-\beta}\bar u (w)\,.
\end{equation}
The critical behavior of the expectation value of the topological
charge density, $q_{{\rm i}\theta}$, can be obtained from that of the
mass gap $m$. Since according to the usual renormalization--group
arguments the free energy per unit volume $F$ is proportional to
$m^2$, one has $q_{{\rm i}\theta} = -{\rm i}\de F / \de\theta
\stackrel{P}{\propto} m\,\de m /\de \theta \stackrel{P}{\propto}
m\,\de m /\de z$. More precisely, writing $m=m_0 {\rm e}^{-\epsilon w}
w^{-\beta}f(w)$, with $f(w)=1+\Oc(\log w / w)$, we have for $y$ (see
Eqs.~\eqref{eq:method1} and \eqref{eq:method1bis}) 
\begin{equation}
  \label{eq:15}
  \begin{aligned}
    y \eqp zq_{{\rm i}\theta} & \stackrel{P}{\propto} m z\f{\de m}{\de z} = 
    -m \f{\de    m}{\de w} \eqp m^2  \left(\epsilon +\f{\beta}{w}
      -\f{\tilde f(w)}{w}\right) \,,
  \end{aligned}
\end{equation}
where $\tilde f(w) = w f'(w)/f(w)= \Oc(\log w/w)$. We can therefore
write
\begin{equation}
  \label{eq:15bla}
  y \eqp  y_0 {\rm e}^{-2\epsilon w}w^{-2\beta}
  \left[1 + \f{\beta}{\epsilon}\f{\log w}{w}
  \right]^{-2\beta}\bar v (w)\,,
\end{equation}
with some constant $y_0$, and with $\bar v(w)= 1 + \Oc(1/w)$. It is
now straightforward to derive expressions for the effective
exponents. They read 
\begin{align}
  \label{eq:ml_yl}
  \begin{aligned}
    \epsilon_m(y) &=    \f{2}{\lambda}\log\f{m_\lambda}{m} 
    \eqp \epsilon \left( 1 +     \f{\beta}{\epsilon}\f{1}{w}
      -   \f{\beta^2}{\epsilon^2}\f{\log
        w}{w^2 + \f{\beta}{\epsilon}w\log w} \right) +\Oc(1/w^2)\\
    & \phantom
    {=\f{2}{\lambda}\log\f{m_\lambda}{m}}=\epsilon \left( 1 +
      \f{\beta}{\epsilon}\f{1}{w+\f{\beta}{\epsilon}\log w } 
    \right) +\Oc(1/w^2)\,,    
  \end{aligned}
  \\
  \label{eq:ml_yl_2}
  \begin{aligned}
    2\epsilon_q(y) &=    \f{2}{\lambda}\log\f{y_\lambda}{y} 
    \eqp 2\epsilon\left( 1 +     \f{\beta}{\epsilon}\f{1}{w}
      -    \f{\beta^2}{\epsilon^2}\f{\log
        w}{w^2 + \f{\beta}{\epsilon}w\log w}
    \right)  +\Oc(1 / w^2)\\
    & \phantom
    {=    \f{2}{\lambda}\log\f{y_\lambda}{y} }
    =2\epsilon \left( 1 +
      \f{\beta}{\epsilon}\f{1}{w+\f{\beta}{\epsilon}\log w } 
    \right) +\Oc(1/w^2)
    \,,
  \end{aligned}
\end{align}
where $w$ has to be traded for $y$ by inverting the following
relation, 
\begin{equation}
  \label{eq:wy}
  \f{1}{2\epsilon}\log\f{y_0}{y} =  w + \f{\beta}{\epsilon}\log w +
  \Oc\left({\log w}/{w}\right)\,.
\end{equation}

\section{Determination of the critical exponents}
\label{sec5}

The presence of the logarithmic factors $w^{-\beta}$ and $w^{-2\beta}$
in Eqs.~\eqref{eq:10} and \eqref{eq:15bla} constitutes a problem for
the numerical analysis. It is well known that the presence of
logarithmic corrections can lead to a wrong estimate of a critical
exponent. In the problem at hand, the main consequences of these
corrections are the $\Oc(1/w)=\Oc(1/\log\f{y_0}{y})$ terms in
Eqs.~\eqref{eq:ml_yl} and \eqref{eq:ml_yl_2}, which lead to rather
large deviations from the value at $y=0$ even for pretty small
$y$. Furthermore, the $\Oc(\log w/w)$ term in Eq.~\eqref{eq:wy}
results into $\Oc(\log\log\f{y_0}{y}/(\log\f{y_0}{y})^2)$ terms in
Eqs.~\eqref{eq:ml_yl} and \eqref{eq:ml_yl_2}, that also give sizeable 
contributions. On top of that, the $\log w$ term in Eq.~\eqref{eq:wy}
spoils the approximate linearity of the relation between
$\log\f{y_0}{y}$ and $w$ at small $w$. As a consequence, these terms
make very difficult to correctly identify the asymptotic value as
$y\to 0$.  

To overcome this problem, it is therefore convenient to first remove
the logarithmic factor, and only after perform the analysis with the
scaling transformations, as suggested in Ref.~\cite{zgz}. An obvious
obstacle is that in principle we do not know the exponent $\beta$. In
Ref.~\cite{zgz} the analysis was performed by taking $\beta=\f{1}{2}$,
in accordance with the theoretical expectation, and trying to
determine the critical exponent by fitting the data for the properly
modified effective exponent obtained from the topological charge. The
results were in agreement with the theoretical expectation. Here we
use another strategy that does not presume any preferred value for
$\beta$: by choosing an arbitrary $\beta$, we obtain two
determinations of the critical exponent by fitting separately the data
for two properly defined effective exponents, involving respectively
the mass gap and the topological charge, {as if} the current value of
$\beta$ were the correct one. We then vary $\beta$, obtaining two sets
of putative critical exponents, one for each observable. The idea is
that for the {correct} choice of $\beta$, the two determinations have
to coincide.  

To determine the mass gap critical exponent from the mass gap data, 
it is therefore convenient to study the behavior of the quantity $\bar
m = m \left(\log\f{1}{z}\right)^{\beta} = m w^\beta$ under the
rescaling $z\to {\rm e}^{\f{\lambda}{2}}z$, or equivalently under the shift
$w\to w - \f{\lambda}{2}$. Analogously, to determine the mass gap
critical exponent from the topological charge data it is convenient to  
consider $\bar y = y (\log \f{1}{z})^{2\beta} = y w^{2\beta}$. To
lowest order\footnote{Due to the resummation done in
  Eq.~\eqref{eq:10}, Eqs.~\eqref{eq:14} and \eqref{eq:14_2}   actually
  contain higher--order terms.} we find from Eqs.~\eqref{eq:ml_yl} and
\eqref{eq:ml_yl_2} 
\begin{align}
  \label{eq:14}
  \f{2}{\lambda}\log\f{\bar m_\lambda}{\bar m} & \eqp 
  \epsilon \left( 1 - 
    \f{\beta^2}{\epsilon^2}\f{\log
      w}{w^2 + \f{\beta}{\epsilon}w\log w} \right) +\Oc(1/w^2)\,,\\
  \label{eq:14_2}
  \f{2}{\lambda}\log\f{\bar y_\lambda}{\bar y} 
  &\eqp 2\epsilon\left( 1 -
    \f{\beta^2}{\epsilon^2}\f{\log
      w}{w^2 + \f{\beta}{\epsilon}w\log w}
  \right)  +\Oc(1 / w^2)\,.
\end{align}
Finally, since to lowest order\footnote{Notice the absence of
  $\Oc(\log\log(y_0/\bar y))$ corrections, which are present in the
  relation between $\log(y_0/y)$ and $w$, see Eq.~\eqref{eq:wy}.}
$w=\log (1/z) = (1/2\epsilon)\log (y_0/\bar y) + o(1)$, with $\bar y_0
= y_0$,  one can write down the relation between the effective
exponents and $\bar y$. 

A possible practical definition of $\bar m$ and $\bar y$ is (recall
that $z=\cosh\f{\vartheta}{2}$)
\begin{equation}
  \label{eq:19}
  \bar m \equiv m \ell^\beta\,,\quad
  \bar y \equiv y \ell^{2\beta}\,,\quad
  \ell = \log\left( 1 + \f{1}{z} \right)  \,.
\end{equation}
However, to avoid distortions at large $\vartheta$ which could worsen
the quality of the numerical analysis, it is preferable to work
instead with the quantities 
\begin{equation}
  \label{eq:20}
  \tilde m \equiv m \left(\f{\ell z}{\log 2}\right)^\beta
  \,, \quad
  \tilde y \equiv y \left(\f{\ell z}{\log 2}\right)^{2\beta}
  \,,
\end{equation}
where we have also introduced a factor $\log 2$ to give $1$ in front 
of $m$ and $y$ at $\vartheta=0$. These quantities are easily seen to
satisfy  
\begin{align}
  \label{eq:21}
  \f{2}{\lambda}\log\f{\tilde m_\lambda}{\tilde m} -\beta &\eqp \epsilon
  \left(1 - 
    \f{\beta^2}{\epsilon^2}\f{\log
      w}{w^2 + \f{\beta}{\epsilon}w\log w}\right)
  +\Oc(1 / w^2)
  \,,\\
  \label{eq:21_2}
  \f{2}{\lambda}\log\f{\tilde y_\lambda}{\tilde y} - 2\beta&\eqp 2\epsilon
  \left(1 - 
    \f{\beta^2}{\epsilon^2}\f{\log
      w}{w^2 + \f{\beta}{\epsilon}w\log w}\right)
  +\Oc(1 / w^2)\,.
\end{align}
For our purposes it is convenient to re--express the quantities on the
l.h.s. of Eqs.~\eqref{eq:21} and \eqref{eq:21_2} as functions of
$\tilde y$. A simple calculation shows that
\begin{equation}
  \label{eq:21chat}
  \log\f{\tilde y_0}{\tilde y} = 2(\epsilon+\beta)w +
  \Oc(\log w/w)\,, 
\end{equation}
where $\tilde y_0 = y_0/(\log 2)^{2\beta}$, which allows to recast
Eqs.~\eqref{eq:21} and \eqref{eq:21_2} as 
\begin{align}
  \label{eq:22}
  \tilde\epsilon_m(\tilde y) &\equiv    \f{2}{\lambda}\log\f{\tilde
    m_\lambda}{\tilde m} 
  -\beta \eqp 
  \epsilon \,{\cal E}
  \left(\textstyle\f{1}{2(\epsilon+\beta)}
    \log\f{\tilde y_0}{\tilde y}\right)
  +\Oc\left(\textstyle(\log\f{\tilde y_0}{\tilde y})^{-2}\right)
  \,,\\
  \label{eq:22_2}
  2\tilde\epsilon_q(\tilde y) &\equiv    \f{2}{\lambda}\log\f{\tilde
    y_\lambda}{\tilde 
    y} - 2\beta \eqp 
  2\epsilon\, {\cal E}
  \left(\textstyle\f{1}{2(\epsilon+\beta)}\log\f{\tilde y_0}{\tilde
      y}\right) 
  +\Oc\left(\textstyle(\log\f{\tilde y_0}{\tilde y})^{-2}\right)
  \,,
\end{align}
where
\begin{equation}
  \label{eq:23}
  {\cal E}(x) = 1 - \f{\beta^2}{\epsilon^2}\f{\log x}{x^2 +
    \f{\beta}{\epsilon}\, x\log x}\,.
\end{equation}
These expressions can be used to fit the numerical data for small
enough $\tilde y$. Since these are low--order approximations to the  
exact expressions, one is introducing a systematic error through the
truncation. We remind the reader that by ``exact'' we mean here up to 
terms originating from powers of $z$ in Eq.~\eqref{eq:6}, which should 
be negligible compared to the logarithmic terms. We mention here
that the values of $\vartheta$ at which we performed the simulations
were chosen in such a way that corresponding pairs of $z$ and
${\rm e}^{\f{\lambda}{2}}z$ could be constructed with $\lambda=0.5$, so
that we did not need any interpolation to compute $\tilde\epsilon_m$
and $\tilde\epsilon_q$.

A practical way to estimate the systematic error due to truncation on
our determinations of the critical exponent is to employ the technique
of constrained fits~\cite{bayes}. This basically consists in adding
more and more subleading corrections to Eqs.~\eqref{eq:22} and
\eqref{eq:22_2}, constraining the corresponding coefficients according
to the available information. When the error on the parameters given
by the fitter settles against increase of the number of terms, it
includes also the contribution of the systematic error due to the
truncation of the exact expression~\cite{bayes}. One can show that by
including higher--order terms, Eqs.~\eqref{eq:22} and \eqref{eq:22_2}
become\footnote{Notice that similar expansions for $\epsilon_m$ and
  $\epsilon_q$ as functions of
  $\Lambda_0=\f{1}{2\epsilon}\log\f{y_0}{y}$ contain, besides a 
  $1/\Lambda_0$ term, also terms proportional to
  $(\log\Lambda_0)^j/\Lambda_0^{j+1}$, which are absent in 
  $\tilde\epsilon_m$ and $\tilde\epsilon_q$.}
\begin{align}
  \label{eq:22bis}
  \tilde\epsilon_m & \eqp 
    \epsilon \,{\cal E}(\Lambda)
    + \sum_{k=2}^\infty\sum_{j=0}^{k-2} h_{jk}^{(m)}
    \f{(\log      \Lambda)^j}{\Lambda^k} 
    \,,\\
    \label{eq:22bis_2}
    2\tilde\epsilon_q  & \eqp 
    2\epsilon \,{\cal E}(\Lambda)    + \sum_{k=2}^\infty\sum_{j=0}^{k-2}
    h_{jk}^{(q)} 
    \f{(\log\Lambda)^j}{\Lambda^k} 
    \,,
\end{align}
where we set $\Lambda \equiv\f{1}{2(\epsilon+\beta)}\log\f{\tilde
  y_0}{\tilde y}$. The constraints on the parameters (``priors'') are
needed to ensure the stability of fits with a rather large number of
parameters. The priors were chosen to be as loose as possible while
leading to fits of good quality. 

\begin{figure}[t]
  \centering
  \includegraphics[width=0.9\textwidth]{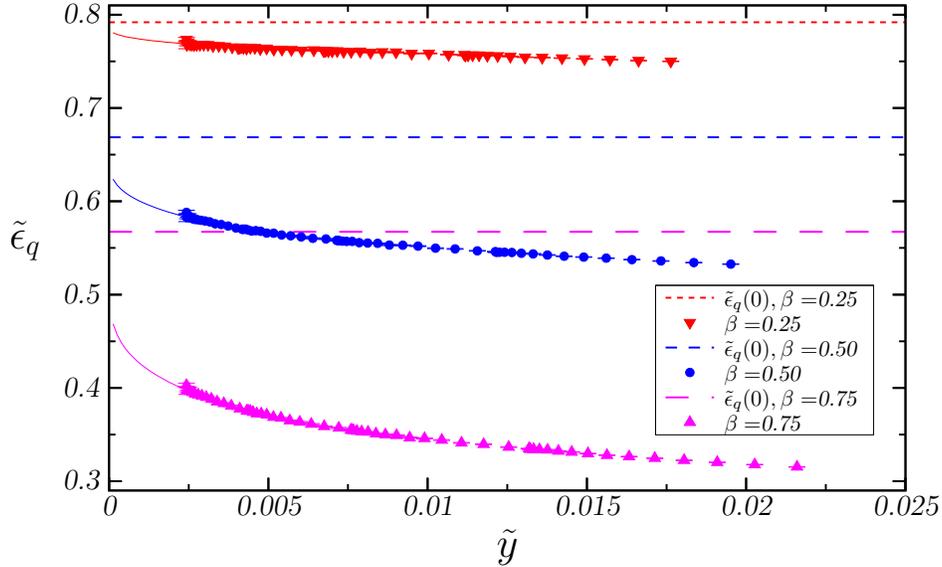}
  \caption{Data (points), fit (solid line), and value at $\tilde y=0$
    (dashed line) for the effective exponent $\tilde\epsilon_q$, for
    three assumed values of $\beta$.} 
  \label{fig:3}
\end{figure}

We have applied this technique to the critical exponent measured from 
the expectation value of the topological charge. In practice we
assumed that the fit parameters obey a Gaussian distribution, with
mean and standard deviation as reported in Tab.~\ref{tab:1}. We used
data up to $\tilde y = 0.01$, and up to 8 fit parameters. The results
of the fit are shown in Fig.~\ref{fig:3}.

\begin{figure}[t]
  \centering
  \includegraphics[width=0.9\textwidth]{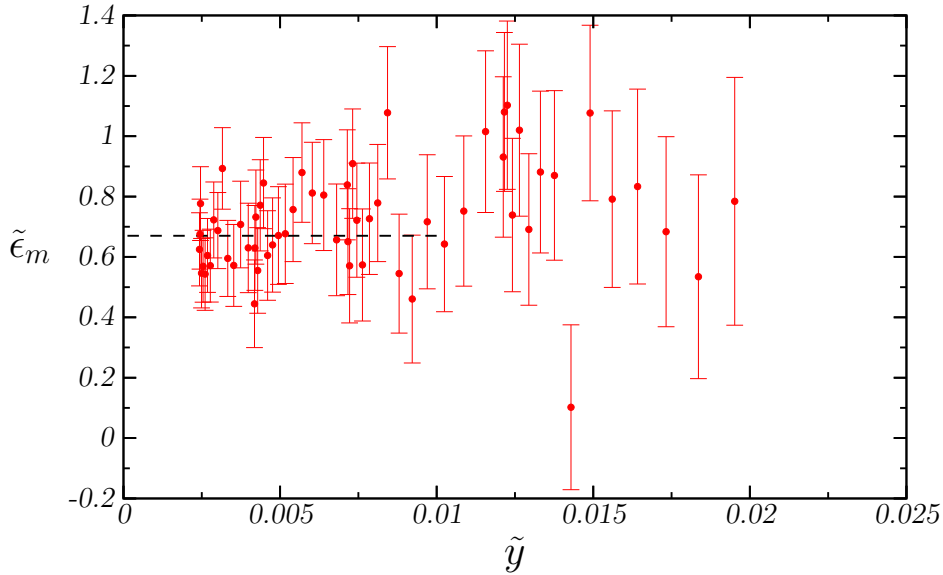}
  \caption{Data for the effective exponent $\tilde\epsilon_m$ (points)
    and result of a fit with a constant (dashed line), assuming
    $\beta=0.5$.}  
  \label{fig:4}
\end{figure}

The same kind of analysis should be performed for the critical
exponent obtained from the mass gap, {\it i.e.}, one should fix
$\tilde y_0$ to the value obtained using the total topological charge
data, and fit the mass gap data including more and more terms in the
expansion to determine the systematic error. However, the quality of
the data for $\tilde\epsilon_m$ is rather poor compared to the very
precise topological charge data, and very hard to improve (we remind
the reader that we made 2 million measurements for each
$\vartheta$). The mass gap data show no clear structure, being
essentially constant within the statistical errors, see
Fig.~\ref{fig:4}. An attempt at including the main contribution and 
the first subleading term in Eq.~\eqref{eq:22bis} results in fits that 
are very sensitive to the choice of priors, indicating that the data
are not good enough for a sophisticated analysis like the one carried
out for the topological charge. However, if the absence of a clear
structure in the data for $\tilde\epsilon_m$ indicates that the size
of the corrections to the value at $\tilde y=0$ is of the same order
of the statistical errors, then a fit to the data with a simple
constant will result into a reasonable estimate of the critical
exponent, and the statistical fluctuations around the central value
will give a reasonable estimate of the error. We shall follow this
latter strategy to determine the critical exponent of the mass gap.

\begin{figure}[t]
  \centering
  \includegraphics[width=0.9\textwidth]{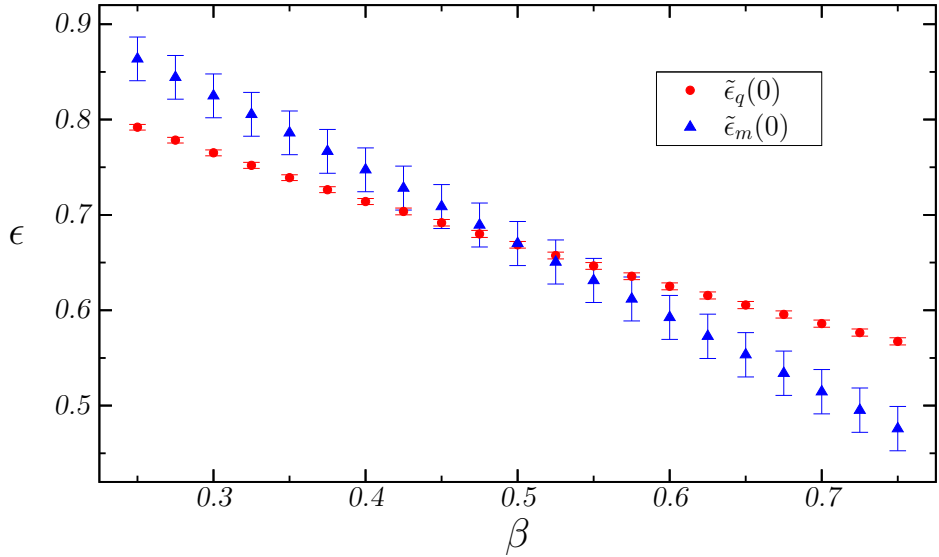}
  \caption{The two determinations $\tilde\epsilon_q(0)$ and 
    $\tilde\epsilon_m(0)$ of the mass gap critical exponent
    $\epsilon$, as a function of the assumed value of $\beta$.} 
  \label{fig:5}
\end{figure}

The results for the two determinations of the critical exponent are
reported in Tab.~\ref{tab:2}. In Fig.~\ref{fig:5} we
compare the two determinations, which clearly cross close to
$\beta=0.5$. We take this value for $\beta$, and for the corresponding
error we take the half--length of the interval $[0.425,0.575]$ where
the two determinations are compatible within one standard deviation,
which results in $\beta=0.50(7)$. For the critical exponent, we take
the average of the values of $\tilde\epsilon_m(0)$ and
$\tilde\epsilon_q(0)$ at $\beta=0.5$, and we quote as error the
half--variation of $\tilde\epsilon_m(0)$ in the range 
$\beta\in [0.425,0.575]$, which yields $\epsilon=0.67(6)$. 
These values are in very good agreement with the theoretical
expectation $\epsilon_{_{\rm WZNW}}=\f{2}{3}$ and $\beta_{_{\rm
    WZNW}}=\f{1}{2}$ for the critical exponent and the exponent of the
logarithmic correction. For completeness, we finish by noting that had
we established the value of $\beta=1/2$ from the very beginning, as in
Ref.~\cite{zgz}, then the determination of $\epsilon$ obtained from
the topological charge would have read
$\tilde\epsilon_q(0)=0.6687_{-0.0036}^{+0.0035}$. 

\section{Conclusions}
\label{sec6}

The present paper deals with the Haldane's conjecture, which states
that the mass gap in the 2D O(3) non--linear sigma model with a
$\theta$--term must vanish as $\theta$ approaches the value $\pi$
according to the precise law given in (\ref{massWZNW}). The aim of the
work is to extract the critical exponent $\epsilon$ ruling the
dominant, power--law behavior of the mass gap near $\theta=\pi$ and
also the elusive exponent $\beta$ of its logarithmic correction,
without any {\it a priori} assumption about their values.

The sign problem hindering the numerical study of the model in the
presence of a non--zero $\theta$ has been circumvented by performing 
Monte Carlo simulations at {imaginary} values of $\theta$ (where
the Euclidean action is real and a positive Boltzmann weight can be
safely defined) and extrapolating the results to real values of $\theta$. 

The basic technique adopted to carry out this extrapolation is that of
scaling transformations proposed in Ref.~\cite{ADGV1}. Had we limited
our analysis to the mass gap only, the target would have been missed,
even in spite of high--statistics Monte Carlo simulations, due to the
intrinsically bad signal--to--noise ratio of this observable (this 
problem arose in Ref.~\cite{alles1}). 

The breakthrough comes by the inclusion in the analysis of a second
observable, the topological charge, for which very accurate
determinations at imaginary $\theta$ can be obtained. Indeed, when the 
compatibility between the extrapolations towards $\theta=\pi$ of the
mass gap and of the topological charge is imposed, a determination of
both the exponents $\epsilon$ and $\beta$ gets within reach, nicely
agreeing with the theoretical prediction.

These determinations, schematically summarized in Fig.~\ref{fig:5},
are $\epsilon=0.67(6)$ and $\beta=0.50(7)$, both in concordance with
the renormalization group prediction of Refs.~\cite{affleck4,affleck5}
shown in (\ref{massWZNW}), namely $\epsilon_{_{\rm WZNW}}=\frac{2}{3}$
and $\beta_{_{\rm WZNW}}=\frac{1}{2}$. 

\section*{Acknowledgements}

Part of the simulations have been run at the computer facility in the
National Laboratories of Gran Sasso and part at CINECA in Bologna,
both in Italy (the latter under the project ``IsC09\_RAFSOSMT (PI)'').  
It is a pleasure to thank the staff of the two computer centers for
their competence and constant help. MG wants to thank V.~Azcoiti,
G.~Di Carlo, E.~Follana and A.~Vaquero for many useful discussions.  
MG is supported by the Hungarian Academy of Sciences under
``Lend\"ulet'' grant No. LP2011--011. This work has been partially
supported by the INFN SUMA project. 

\begin{table}[t]
  \centering
  \begin{tabular}[h]{c|cc}
    parameter & mean & standard deviation \\ \hline
          $\epsilon$ & 1 & 100 \\
          $\tilde y_0$ & 1.0 & 1.0 \\
          $h_{02}^{(q)}$ & 0.1 & 0.1 \\
          $h_{13}^{(q)}$ & -0.1 & 0.1 \\
          $h_{03}^{(q)}$ & 0.1 & 0.1 \\
          $h_{24}^{(q)}$ & 0.0 & 0.1 \\
          $h_{14}^{(q)}$ & 0.0 & 0.01 \\
          $h_{04}^{(q)}$ & 0.0 & 0.01
  \end{tabular}
  \caption{Priors used in the constrained fits for
    $\tilde\epsilon_q(\tilde y)$.} 
  \label{tab:1}
\end{table}

\begin{table}[thb]
  \centering
  \begin{tabular}[h]{c|cc}
$\beta$ & $\tilde\epsilon_q(0)$ & $\tilde\epsilon_m(0)$ 
\\ \hline 
0.250 &  $0.7919_{ -0.0030}^{+ 0.0028} $ & $0.864\pm 0.023$ \\
0.275 &  $0.7784_{ -0.0030}^{+ 0.0029} $ & $0.844\pm 0.023$ \\
0.300 &  $0.7651_{ -0.0031}^{+ 0.0030} $ & $0.825\pm 0.023$ \\
0.325 &  $0.7520_{ -0.0031}^{+ 0.0030} $ & $0.805\pm 0.023$ \\
0.350 &  $0.7391_{ -0.0032}^{+ 0.0031} $ & $0.786\pm 0.023$ \\
0.375 &  $0.7264_{ -0.0032}^{+ 0.0031} $ & $0.767\pm 0.023$ \\
0.400 &  $0.7140_{ -0.0032}^{+ 0.0031} $ & $0.747\pm 0.023$ \\
0.425 &  $0.7037_{ -0.0036}^{+ 0.0035} $ & $0.728\pm 0.023$ \\
0.450 &  $0.6918_{ -0.0036}^{+ 0.0035} $ & $0.709\pm 0.023$ \\
0.475 &  $0.6801_{ -0.0036}^{+ 0.0035} $ & $0.689\pm 0.023$ \\
0.500 &  $0.6687_{ -0.0036}^{+ 0.0035} $ & $0.670\pm 0.023$ \\
0.525 &  $0.6574_{ -0.0036}^{+ 0.0035} $ & $0.651\pm 0.023$ \\
0.550 &  $0.6464_{ -0.0036}^{+ 0.0035} $ & $0.631\pm 0.023$ \\
0.575 &  $0.6357_{ -0.0036}^{+ 0.0035} $ & $0.612\pm 0.023$ \\
0.600 &  $0.6251_{ -0.0036}^{+ 0.0035} $ & $0.592\pm 0.023$ \\
0.625 &  $0.6156_{ -0.0037}^{+ 0.0037} $ & $0.573\pm 0.023$ \\
0.650 &  $0.6055_{ -0.0037}^{+ 0.0037} $ & $0.553\pm 0.023$ \\
0.675 &  $0.5957_{ -0.0038}^{+ 0.0037} $ & $0.534\pm 0.023$ \\
0.700 &  $0.5860_{ -0.0038}^{+ 0.0037} $ & $0.515\pm 0.023$ \\
0.725 &  $0.5766_{ -0.0038}^{+ 0.0038} $ & $0.495\pm 0.023$ \\
0.750 &  $0.5674_{ -0.0038}^{+ 0.0038} $ & $0.476\pm 0.023$ 
  \end{tabular}
  \caption{Results for $\tilde\epsilon_q(0)$, obtained with a
    8--parameter constrained fit of $\tilde\epsilon_q$, and results
    for $\tilde\epsilon_m(0)$ obtained with a fit of
    $\tilde\epsilon_m$ with a constant, for several assumed values of  
    $\beta$.} 
  \label{tab:2}
\end{table}


\clearpage

\begin{thebibliography}{99}

\bibitem{haldane1} F. D. M. Haldane, Phys. Lett. {\bf 93A}, 464
  (1993). 

\bibitem{haldane2} F. D. M. Haldane, Phys. Rev. Lett. {\bf 50}, 1153
  (1983). 

\bibitem{haldane3} F. D. M. Haldane, J. Appl. Phys. {\bf 57}, 33
  (1985). 

\bibitem{lieb} E. H. Lieb, T. Schultz and D. Mattis, Ann. Phys. {\bf
    16}, 407 (1961). 

\bibitem{botet} R. Botet, R. Jullien and M. Kolb, Phys. Rev. {\bf
    B30}, 215 (1984). 

\bibitem{affleck1} I. Affleck, E. H. Lieb, Lett. Math. Phys. {\bf 12},
  57 (1986). 

\bibitem{affleck2} I. Affleck, T. Kennedy, E. H. Lieb and H. Tasaki,
  Phys. Rev. Lett. {\bf 59}, 799 (1987). 

\bibitem{affleck3} I. Affleck, T. Kennedy, E. H. Lieb and H. Tasaki,
  Commun. Math. Phys. {\bf 115}, 477 (1988).

\bibitem{schollwock} U. Schollw\"ock and T. Jolicoeur,
  Europhys. Lett. {\bf 30}, 493 (1995). 

\bibitem{Affleck:1991tj}
  I.~Affleck,
  Phys.\ Rev.\ Lett.\  {\bf 66}, 2429 (1991).

\bibitem{fradkin} E. Fradkin, ``Field theories of condensed matter
  systems'', Addison--Wesley Pub. Company, Redwood city (1991). 

\bibitem{torrero} C. Torrero, O. Borisenko, V. Kushnir, B. All\'es and
  A. Papa, PoS LATTICE {\bf 2013}, 338 (2013).

\bibitem{bietenholz} W. Bietenholz, A. Pochinsky and U.--J. Wiese,
  Phys. Rev. Lett. {\bf 75}, 4524 (1995). 

\bibitem{BNPW} M.~B\"ogli, F.~Niedermayer, M.~Pepe and U.--J.~Wiese,
  JHEP {\bf 1204}, 117 (2012).

\bibitem{dFPW} P.~de Forcrand, M.~Pepe and U.--J.~Wiese,
  Phys.\ Rev.\ {\bf D86}, 075006 (2012).

\bibitem{alles1} B. All\'es and A. Papa, Phys. Rev. {\bf D77}, 056008 
  (2008). 

\bibitem{azcoiti1} V. Azcoiti, G. Di Carlo and A. Galante,
  Phys. Rev. Lett. {\bf 98}, 257203 (2007). 

\bibitem{zgz} V.~Azcoiti, G.~Di Carlo, E.~Follana and M.~Giordano,
  Phys.\ Rev.\ {\bf D86}, 096009 (2012).

\bibitem{affleck4} I. Affleck and F. D. M. Haldane, Phys. Rev. {\bf
    B36}, 5291 (1987). 

\bibitem{affleck5} I. Affleck, D. Gepner, H. J. Schulz and T. Ziman,
  J. Phys. A: Math. Gen. {\bf 22}, 511 (1989). 

\bibitem{wess} J. Wess and B. Zumino, Phys. Lett. {\bf 37B}, 95
  (1971). 

\bibitem{novikov} S. P. Novikov, Sov. Math. Dokl. {\bf 24}, 222
  (1981). 

\bibitem{witten} E. Witten, Commun. Math. Phys. {\bf 92}, 455 (1984). 

\bibitem{pruisken} A. M. M. Pruisken and I. S. Burmistrov,
  Ann. Phys. {\bf 316}, 285 (2005). 

\bibitem{belavin} A. A. Belavin and A. M. Polyakov, JETP Lett. {\bf
    22}, 245 (1975). 

\bibitem{hasenfratz} P. Hasenfratz, M. Maggiore and F. Niedermayer,
  Phys. Lett. {\bf B245}, 522 (1990). 

\bibitem{controzzi} D. Controzzi and G. Mussardo,
  Phys. Rev. Lett. {\bf 92}, 021601 (2004). 

\bibitem{alles2} B. All\'es, G. Cella, M. Dilaver 
  and Y. G\"und\"u\c c, Phys. Rev. {\bf D59}, 067703 (1999).  

\bibitem{berg} B. Berg and M. L\"uscher, Nucl. Phys. {\bf B190}, 412 (1981).

\bibitem{alles3} B. All\'es, A.~Buonanno and G. Cella,
  Nucl. Phys. {\bf B500}, 513 (1997). 

\bibitem{ADGV1}
  V.~Azcoiti, G.~Di Carlo, A.~Galante and V.~Laliena,
  Phys.\ Lett.\  {\bf B563}, 117 (2003).

\bibitem{ADGV2}
  V.~Azcoiti, G.~Di Carlo, A.~Galante and V.~Laliena,
  Phys.\ Rev.\  {\bf D69}, 056006 (2004).

\bibitem{AFV}
  V.~Azcoiti, E.~Follana and A.~Vaquero,
  Nucl.\ Phys.\  {\bf B851}, 420 (2011).

\bibitem{BDSL} G.~Bhanot, R.F.~Dashen,
  N.~Seiberg and H.~Levine, Phys. Rev. Lett. {\bf 53}, 519 (1984).

\bibitem{Nogradi}
  D.~N\'ogr\'adi,
  JHEP {\bf 1205}, 089 (2012).

\bibitem{bayes} G.~P.~Lepage, B.~Clark, C.~T.~H.~Davies,
  K.~Hornbostel, P.~B.~Mackenzie, C.~Morningstar and H.~Trottier, 
  Nucl.\ Phys.\ Proc.\ Suppl.\  {\bf 106}, 12 (2002).
  

\end{thebibliography}
\end{document}